# Transmission Kikuchi diffraction mapping induces structural damage in atom probe specimens

Baptiste Gault[1,2], Heena Khanchandani[1], T.S. Prithiv[1], Stoichko Antonov[1,3], T. Ben Britton[4]

[1]*Department of Microstructure Physics and Alloy Design, Max-Planck-Institut für Eisenforschung, Max-Planck-Str. 1, 40237 Düsseldorf, Germany.*

[2]*Department of Materials, Royal School of Mines, Imperial College, Prince Consort Road, London SW7 2BP, United Kingdom.*

[3]*National Energy Technology Laboratory, 1450 Queen Ave. SW, Albany, OR 97321, United States.*

[4]*Department of Materials Engineering, University of British Columbia, Frank Forward Building, 309-6350 Stores Road, Vancouver, BC, Canada V6T 1Z4*

## Abstract

Measuring local chemistry of specific crystallographic features by atom probe tomography (APT) is facilitated by using transmission Kikuchi diffraction (TKD) to help position them sufficiently close to the apex of the needle-shaped specimen. However, possible structural damage associated to the energetic electrons used to perform TKD is rarely considered and is hence not well-understood. Here, in two case studies, we evidence damage in APT specimens from TKD mapping. First, we analyze a solid solution, metastable β-Ti-12Mo alloy, in which the Mo is expected to be homogenously distributed. Following TKD, APT reveals a planar segregation of Mo amongst other elements. Second, specimens were prepared near Σ3 twin boundaries in a high manganese twinning-induced plasticity steel, and subsequently charged with deuterium gas. Beyond a similar planar segregation, voids containing a high concentration of deuterium, i.e. bubbles, are detected in the specimen on which TKD was performed. Both examples showcase damage from TKD mapping leading to artefacts in the compositional distribution of solutes. We propose that the structural damage is created by surface species, including H and C, subjected to recoil from incoming energetic electrons during mapping, thereby getting implanted and causing cascades of structural damage in the sample.

## Introduction

The wide spread use of focused ion beam (FIB) combined with scanning electron microscopes (SEM) has been an enabler for site-specific specimen preparation for atom probe tomography (APT) (Miller et al., 2005; Larson et al., 1998; Prosa & Larson, 2017). Analyses of grain and phase boundaries have been reported in a wide range of engineering materials including steels (van Landeghem et al., 2017; Danoix et al., 2016; Moszner et al., 2014), aluminum alloys (Zhao et al., 2020) and titanium alloys (Yan et al., 2019).

Even using site specific FIB lift out, it can still be challenging to position a grain boundary at the apex of an APT specimen and to be able to capture it before the specimen fractures. In recent years, it has been proposed to perform transmission Kikuchi diffraction (TKD) on the needle-shaped APT specimens, to facilitate the analysis by APT of specific crystallographic features, particularly grain boundaries (K Babinsky et al., 2014). TKD is the analysis of patterns similar to those obtained from EBSD, however the signal originates from electrons transmitted through the specimen and emitted from the backside of the illuminated specimen (Trimby, 2012; Keller & Geiss, 2012). As the sample is thinner, there is relatively less scattering and so TKD enables high-resolution crystallographic analysis with a potential spatial resolution down to the range of 10 nm.

Provided that the SEM/FIB is equipped with an 'off axis' electron backscatter diffraction (EBSD) detector, TKD can be performed over the course of the preparation of APT specimens, while the specimen is aligned with the ion column which enables to alternate TKD analysis and annular milling without moving and realigning the sample, as depicted in the schematic in Figure 1a. This makes it particularly useful for a crystallographic interface close to the specimen's apex, as conventionally these features are challenging to observe using conventional SEM imaging. As examples, this technique combination has opened up the combination of TKD and APT to enable systematic analysis of the misorientation between adjacent grains/phases and study the composition of grain and twin boundaries in a range of metallic and semiconducting materials (K Babinsky et al., 2014; Breen et al., 2017; Schwarz et al., 2018; Tsai et al., 2021; Prithiv et al., 2022).

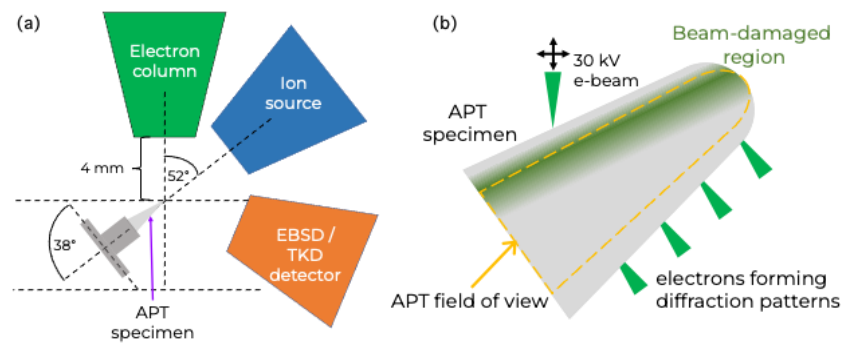

*Figure 1. (a) schematic of the transmission Kikuchi diffraction setup; (b) detail*

In all characterization approaches, damage during sample preparation can significantly alter the interpretation of data. Knowledge of different effects can be informed from the extensive studies of cascades of damage caused by energetic neutrons, protons or ions in nuclear materials work (e.g. Nordlund et al., 2018). The formation and accumulation of defects from these cascades of defects can evolve over time, and vacancies can cluster for instance and self-interstitials can condensate to form dislocation loops. The location of damage, and the accumulation of defects are constrained by the particle matter interactions, especially as the high energy incident particle slows down and its interaction cross-sections with atoms from the target becomes sufficient to cause the first event in the damage cascade. This means that damage tends to be found near surface, where the distance is related to the incident ion type and energy, and the nature of the sample. For example, in FIB studies, ion damage can be induced by the incident high energy Ga+ ions that create point defects and the gallium can implant into the material (Thompson et al., 2007a). These defects are well understood, and APT was one of the techniques used to compare, and contrast electro-polished and FIB-prepared specimens to create FIB-based sample preparation routes that reduced these artifacts (Felfer et al., 2012).

In contrast, electron damage is more rarely considered in FIB-SEM preparation, especially for 'low energy' electrons used in a scanning electron microscope (typically up to 30 kV). However, it is has been known for over 50 years that electron beam damage may also include the creation of lattice vacancies and interstitials, through the recoil of knocked-off atoms from the sample induced by the incoming high-energy electrons, which can, in turn, initiate cascades similar to those produced by e.g. ions, as reviewed by Vajda already in the 1970s (Vajda, 1977). Most studies of damage related to electron irradiation were for a much higher energy range, up to MeV (Hautojärvi et al., 1979; Vehanen et al., 1982) and often from indirect measurements such as positron annihilation or resistivity. For thin samples (thickness in the range of approx. 100 nm) and acceleration voltages typically used for transmission-electron microscopy (TEM), i.e. 80–300kV, evidence of damage has been reported (Fujimoto & Fujita, 1972; Egerton, 2019a), particularly in the case of samples imaged by cryogenic-TEM (Baker & Rubinstein, 2010), which are often biological or soft materials. However, it is often to be limited as the cross section of interaction between electrons and the atoms from the sample itself is lower at higher energies and only when electrons are slowed down by successive inelastic scattering events as electrons are travelling through the material.

As TKD is a SEM based technique, it is performed at lower voltages, typically 20–30 kV and currents of up to a few nA (K. Babinsky et al., 2014; Rice et al., 2016). This produces TKD patterns that are similar to conventional EBSD patterns and therefore easy to analyze with prior EBSD-based software.In this energy range beam damage is often considered non-existant (Egerton et al., 2004). However, there are reports of vacancy generation in semiconductors (Nykänen et al., 2012) and diamond (Schwartz et al., 2012). In metals, Gu et al. reported possible knock-on damage created by beam damage at 30kV of acceleration in a Zr-based alloy (Gu et al., 2017). Yet details of the mechanisms are rather unclear, and Egerton (Egerton, 2019b) suggests that, in metals, the threshold energy to cause substantial damage is in principle not reached by these lower energy incoming electrons (e.g. up to 30 kV)`.

An important aspect for the APT community is that the TKD samples used for needle analysis are different to many of the TEM-like samples used in other studies. The needle is a curved surface that can be several hundred of nanometers in thickness, which represents a very small volume in comparison to a conventional specimen for SEM. The incoming electron beam is also at a high angle with respect to the normal to the surface, which can enhance the risk of structural damage (Egerton et al., 2004). Damage from the incoming electron beam could hence potentially be extending beyond just the surface and into the bulk of the specimen, as illustrated in Figure 1b.

In a recent study, unexpected compositional fluctuations in a twinning-induced plasticity steel analyzed by APT were suspected to be associated to damage from TKD, as reported in a previous study (Khanchandani et al., 2021). This has motivated the present work, wherewe performed a targeted study to ascertain the nature of observed artefacts within APT analysis of specimens following TKD compared to similar specimens not mapped by TKD. To help provide a systematic and pragmatic understanding of this issue, two different materials systems were selected. In the first case, an artificial compositional variation in the form of a segregation of light and heavy species in a planar feature as a constant depth below the specimen's surface is revealed in a metastable β-Ti alloy following exposure to electrons during TKD. In the second case, a coherent Σ3 twin boundary in a high manganese twinning induced plasticity (TWIP) steel was targeted. These boundaries should show no elemental segregation due to their low energy (Marceau et al., 2013; Herbig et al., 2014). Following TKD mapping, we charged the APT specimens with deuterium to examine its segregation to the boundary. In this second case, a similar artefact is observed, along with void or bubble formation. Our results are rationalized based on the expected damage induced in the specimen during illumination by electrons during TKD mapping.

## Materials and methods

For the first case study, the material was a Ti-12Mo-based (wt.%) alloy, produced by vacuum arc melting from high purity feedstock. The produced ingot was cold rolled to ~80% height reduction and recrystallized at 900°C for 30 minutes.

For the second case study, we selected a Fe-26.9Mn-0.28C (wt.%) TWIP steel that was synthesized by strip casting (Daamen et al., 2013a) and homogenized by an annealing treatment carried out at 1150°C for 2 hours (Daamen et al., 2013b), and studied in details in Ref.(Khanchandani et al., 2023). Subsequently, the TWIP steel sample was cold-rolled to achieve a 50% thickness reduction, and subjected to a recrystallization anneal at 800°C for 20 minutes, followed by water cooling to room temperature.

To help select the features of interest for the APT analysis, initial EBSD mapping was performed using a Zeiss Sigma 500 SEM equipped with an EDAX/TSL system with a Hikari camera at an accelerating voltage of 15 kV, a beam current of 9 nA, a scan step size of 0.5 µm, a specimen tilt angle of 70°, and a working distance of 14 mm (Zaefferer & Habler, 2017).

Once grain boundaries were identified, the APT specimens were prepared by using the site-specific lift-out procedure outlined in Ref. (Thompson et al., 2007b) using FEI Helios NanoLab 600i dual-beam FIB/SEM instrument. Hexagonal-grid based TKD mapping was performed on the needles within this microscope using a step size of 20 nm. TKD patterns were captured using an accelerating voltage of 30 kV, a probe current of 2.7 nA and an exposure time of ~0.05 s. EBSD and TKD data analyses were performed using OIM Data Analysis 7.0.1 (EDAX Inc.) software.

APT experiments were conducted on a LEAP 5000 XR instrument (CAMECA Instruments Inc. Madison, WI, USA). TWIP steel specimens were measured in voltage pulsing mode at a set-point temperature of 70 K, 15% pulse fraction, 200 kHz pulse repetition rate, and an average detection rate of 5 ions per 1000 pulses. Ti-12Mo specimens were measured in laser pulsing mode at a set-point temperature of 40 K, with a laser pulse repetition rate of 125 kHz, a pulse energy of 15 pJ, and detection rates of up to 1 ion per 100 pulses.

## Results and discussion

### Case study I

Figure 2a and b are the top and side view, respectively, of the tomographic reconstruction obtained from the APT analysis of a specimen of Ti-12Mo that was not exposed to TKD. Impurities, namely C and Ni atoms are detected in the analysis and displayed. These impurities appear randomly distributed across the dataset, as expected for a solid solution. A second specimen was prepared from the same lifted-out region, within the same crystallographic grain and did not contain any grain boundary or any other notable microstructural feature. It was subjected to TKD mapping prior to analysis, see Suppl. Material Figure S1. TKD mapping was carried out from the apex of the needle to its bottom, and mainly to show its effects on the material by APT, not to search of a particular feature of interest.

Figure 2c–d are top and side views for the APT reconstruction from the analysis of this second specimen. The pink arrows point to regions of anomalous compositional variations, and the electron illumination during TKD mapping was coming from the right side of the reconstructed data. As highlighted by two sets of isosurfaces, Ni and Mo are heterogeneously distributed, and seem to be forming a planar feature. This plane is nearly parallel to the edge of the reconstructed dataset along the z-direction, and appears to start only below a certain depth, even though towards the specimen's apex regions of higher Ni concentration also appeared.

Figure 2e displays a series of detector hit maps containing 4 x $10^6$ detected ions, showing the evolution of the pattern as the specimen is being field evaporated. The pink arrows point to some of the anomalies in the point density that did not appear in the specimen's maps obtained for the specimen not exposed to TKD. On these projected hit maps, the planar feature appears as a series of high density regions forming a curved linear is visible in (iii–iv) and keeps propagating down until (v–vi) in which a less obvious second feature is also visible. Even towards the start of the run, in the first hit maps, anomalies in the point density are visible, see those marked by a pink arrow in Figure 2e-ii for instance. The linear features appear to be moving progressively out of the field-of-view as the specimen gets blunter, which could indicate that it is parallel to the sides of the specimen, as suggested in Figure 1b.

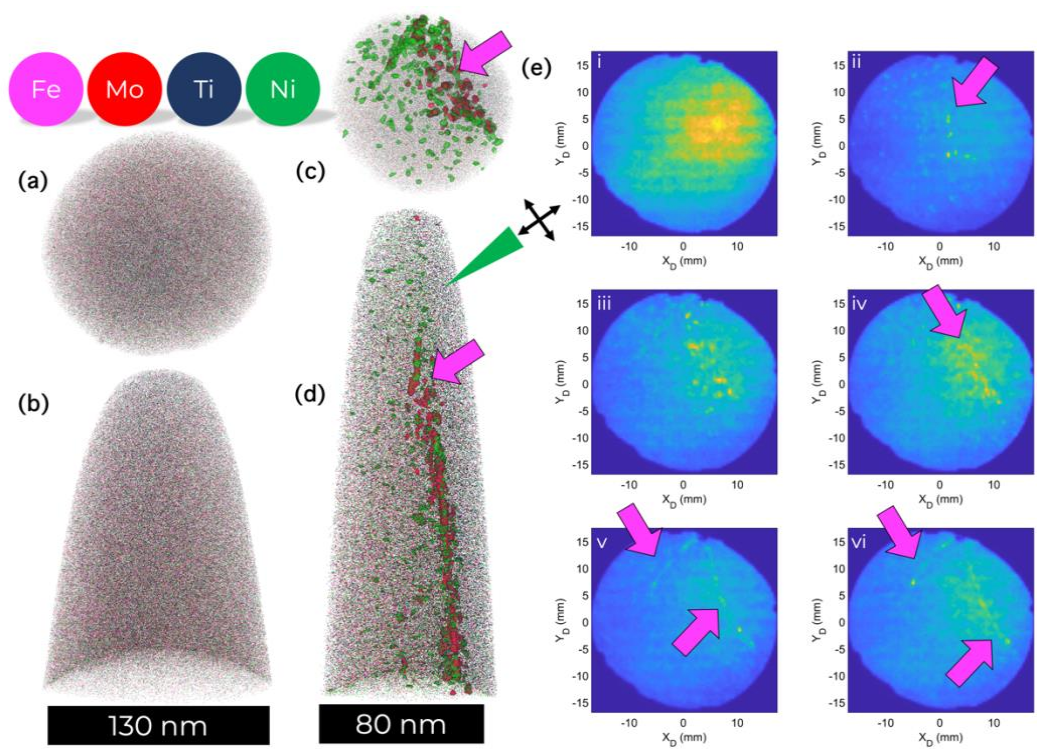

*Figure 2. Top (a) and side (b) view of the 3D APT reconstruction of a specimen not exposed to TKD. Top (c) and side (d) view of the 3D APT reconstruction of a specimen exposed to TKD (only a fraction of the atoms are shown for clarity). The red isosurface encompasses regions containing more than 5 at.nm⁻³ Mo; the green isosurfaces encompass regions containing more than 0.25 at.nm⁻³ of Ni. The green triangle indicates the direction of the incident electron beam. (e) series of successive detector*

The composition across this planar feature, Figure 3a, is quantified by using a 1D composition profile calculated in a cuboidal region of interest positioned normally across over 120nm of the length of the interface, and with a width of approx. 25 nm, in the direction indicated by the pink arrow. This profile showcases a substantial increase in the concentration of elements including C, Ni, O, as well as Mo. In principle, this specimen is devoid of any microstructural feature that could explain the segregation. If Ni was reported to be a fast diffuser in Ti (Lutjering & Williams, 2003), diffusion of Mo at or near room temperature is most unexpected (Bian et al., 2019). The curvature, orientation, and inhomogeneous composition profile suggests that this is an artefact that is related to the TKD mapping. N is also found segregated but not plotted in this 3D rendering for clarity. Overall, the composition of carbon in the sample is almost twice as high in the sample post-TKD (0.11 at % vs. 0.06 at%), and that of N (0.06 at% vs. 0.04 at%) and H (14.9 at% vs. 13.4 at%) is also overall higher. It is anticipated that this high level of H is related to the ingress of H during the preparation of the specimens, but it is noteworthy that is found more highly concentrated in the region high in Mo, and segregated to the interface. There is also in this region a higher fraction of Al3+ and Ti3+, indicative of a higher electric field, which supports that the hydrogen is inside of the specimen and not from the chamber's residuals (Chang et al., 2018).

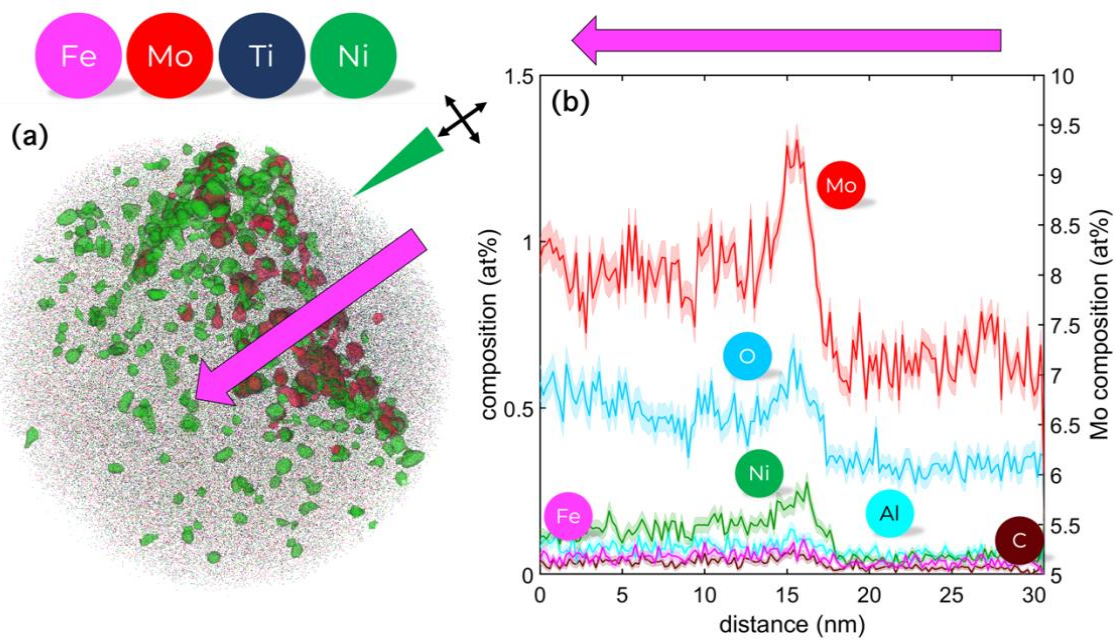

*Figure 3. (a) Top view of the 3D elemental maps with similar isosurfaces as in Figure 2 (c–d). The green triangle indicates the direction of the incident electron beam. (b) 1-D composition profile across the region indicated by the pink arrow.*

To summarize, the observed elemental segregation that is likely correlated with TKD-induced electron beam related damage is not at the top-most surface, but tens of nanometers below, in a region that appears to have a maximum of compositional changes, as illustrated in Figure 1b.

To rationalize these results, let us consider possible mechanisms leading to our observations of the increased composition at this planar feature and below. First, an increase in temperature of only a few K is expected from the illumination by electrons at the current and acceleration voltage used herein (Hobbs, 1990). This cannot explain the observed profile. Second electrons can directly induce the observed defects by knocking off atoms from their lattice sites, particularly as they are being slowed down through successive scattering events with atoms from the lattice, increasing their interaction cross section with specific species but decreasing the probability of causing direct damage.

Third, hydrocarbons are known to get deposited on the sample surface (Egerton & Rossouw, 1976), and moisture from the FIB-vacuum chamber can also deposit on the surface and form oxides and atomic hydrogen, particularly on Ti, as discussed extensively across the literature (Ding & Jones, 2011; Chang et al., 2019). The presence of this carbon-based contamination is in part what motivates the final clean up with a low acceleration voltage (2–5 kV) performed after electron imaging of the specimens, which greatly improves data quality and yield (Herbig & Kumar, 2021). A similar clean-up is done at the end of the FIB milling to remove the regions severely damaged by the ion implantation, yet this step cannot be held responsible for the observations post-TKD mapping since it was also performed on the first specimen, in which no planar feature was imaged.

Therefore, we consider the interaction of incoming electrons and their collisions with C-, O- or H- containing adsorbed surface species, or atoms from the material itself.By virtue of conservation of energy and momentum, these surface-originating ion species would be also be pushed inside the specimen. Depending on the angle of incidence and respective mass, these ion species could acquire a relative high velocity and knock atoms off their lattice site and cause cascades of defects (Vajda, 1977). This is well known in the semiconductor community, as this process underpins the process of recoil implantation by electrons for doping semiconductors (Ito et al., 1978; Wada, 1981; Kozlovski et al., 2015). For this mechanism, the interaction cross-section is expected to be proportional to the atomic number of the atoms (Ito et al., 1978; Wada, 1981; Kozlovski et al., 2015), which could explain why Mo and Ni appear affected (as they have a larger cross section). Furthermore, this process could explain the observed increase in the concentration of carbon inside the specimen subsequent to TKD mapping for instance, as had been reported in the case of ion implantation in (Dagan et al., 2015).

Ultimately, these mechanisms (electron-matter caused damage, surface implantation and associated ion damage) can proceed together and combine to explain the observed damage forming a planar feature where the compositional change is maximum. This could be interpreted as the peak of the induced damage, similar to reports found in many other studies of radiation damage by ions (e.g. Bachhav et al., 2014; Etienne et al., 2010; Meslin et al., 2013; Lloyd et al., 2019). As the geometry of the APT needle is different to these studies, it is reasonable that this (near) planar feature observed in APT moves as the specimen's diameter increases, as suggested in Figure 1 (b), suggesting that the peak of damage is at a certain, nearly constant depth below the surface. The range of depth at which this peak damage occurs appears consistent with H-implantation at 10–30 keV (Bissbort & Becker, 2022). Towards the tip of the sample, it may be that the specimen is too thin for this peak damage to be within the APT field of view as well.

*Case study II*

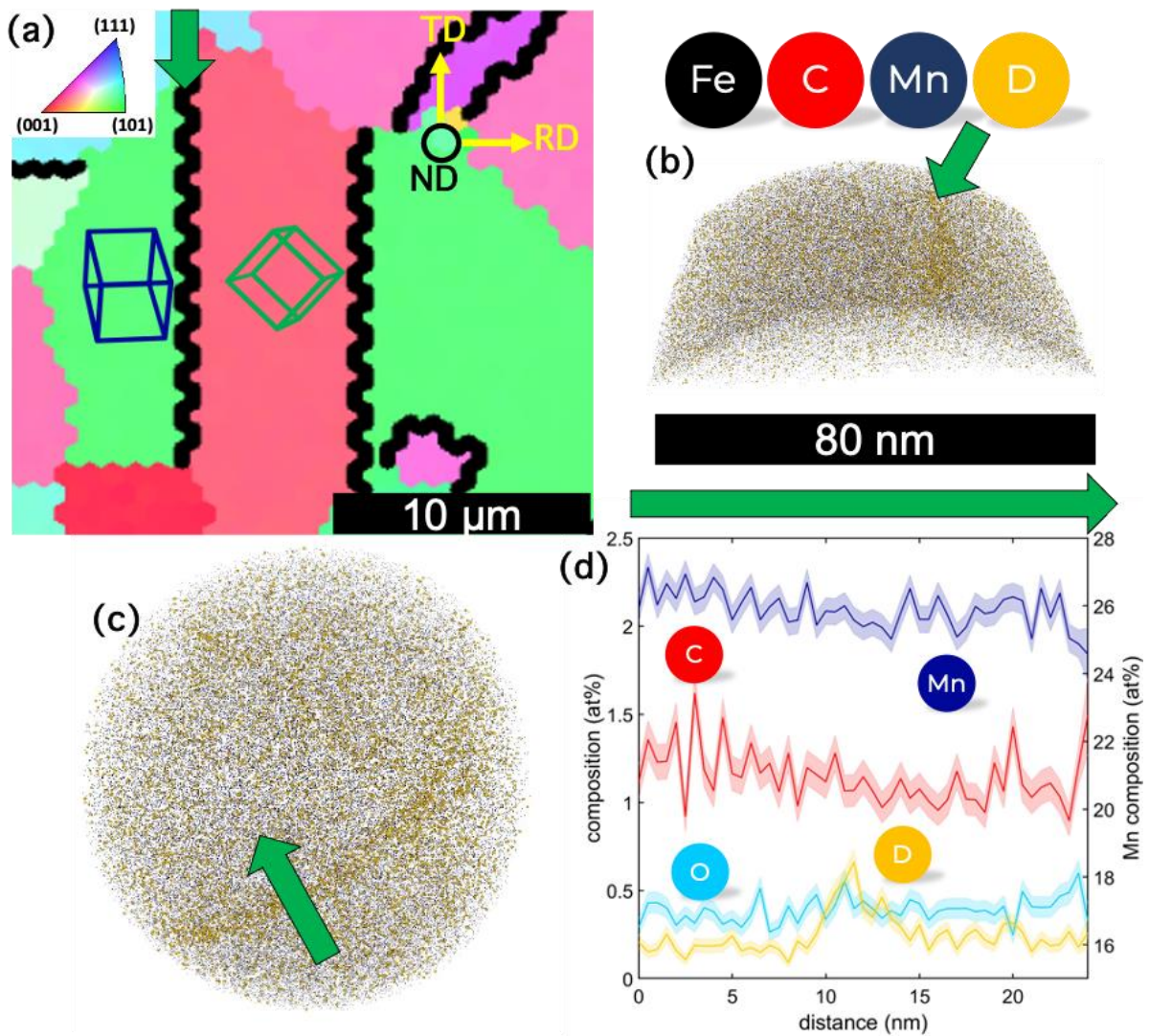

*Figure 4. (a) Electron backscatter diffraction-inverse pole figure (IPF) map, coloured with respect to the normal direction, used to identify the coherent Σ3 twin boundary which was selected for APT analysis; TD – transverse direction; ND – normal direction out of the plane; RD – rolling direction; side (b) and top (c) view of the 3-D elemental map showing iron (Fe), manganese (Mn), carbon (C) and deuterium (D) atoms distribution (only a fraction of the atoms are shown for clarity); (d) One-dimensional composition profile across the interface indicated by green arrow.*

Two APT specimens were prepared from a region near a coherent Σ3 twin boundary of the recrystallized TWIP steel sample which is indicated by the green arrow in the EBSD-inverse pole figure (IPF) map in Figure 4a. TKD was performed on one specimen and not performed on a second specimen. Both specimens were then exposed together to a deuterium gas atmosphere, under a pressure of 250 mbar in the gas charging chamber described in Ref. (Khanchandani et al., 2021a) at room temperature for 6 hours. Room temperature was chosen for introducing deuterium into the APT specimens to avoid any changes in the stacking fault energy of the selected TWIP steel at high temperatures (Saeed-Akbari et al., 2009). Studies also suggest that the shape of APT specimens can change when they are subjected to a heating treatment (Boling et al., 1958). After deuterium charging in the gas charging chamber, both specimens were transferred to the atom probe for analysis in a precooled UHV suitcase, using the infrastructure described in Ref.(Stephenson et al., 2018).

Figure 4b–c are the results of the APT specimen following charging and on which TKD was not performed. The 3-D elemental map in Figure 4b–c shows iron, manganese, carbon and the detected deuterium atoms, and Figure 4c is the top view from the same distribution. The deuterium content obtained from the bulk composition analysis was 0.027 ion%, but the deuterium atoms appear segregated at an interface, which we assume to be the targeted grain boundary. This could not be ascertained due since TKD was not performed on this specimen. The slight decrease in C and Mn that we can observe supports this hypothesis, based on previous reports in a similar alloy of APT analysis of a coherent twin (Herbig et al., 2015). The peak D composition, quantified by a 1-dimensional composition profile obtained along a cuboidal region-of-interest positioned across the interface, is in the range of approx. 0.5 at.%. By comparing the D/H ratio, using the reference data from (Khanchandani et al., 2022) plotted in Suppl. Material Figure S2, we can confirm that the charging had been effective, and the observation of a signal at 2 Da is not only related to an artefact from the APT analysis.

The second specimen was mapped by TKD, and the corresponding unique grain map is plotted in Figure 5a. This shows the presence of a boundary approximately 100 nm below the needle's apex. Figure 5b is the top view of the reconstructed APT analysis showing the distribution of iron, manganese, carbon, and deuterium atoms, whereas Figure 5c is a side view from the same distribution. Two sets of isosurfaces highlight regions of relatively high content in O and D. The left part of the reconstructed data appears enriched in deuterium, with a distribution that is not homogenous. In this case again, there appear to be a curved, planar feature containing a high density of D-rich regions, similar to the two regions observed in Figure 3.

Figure 5d is a series of detector hit maps, similar to those in Figure 2e. Unexpected point density fluctuations, pointed to by the pink arrows, are also imaged in the form of a curved linear feature propagating down as the specimen is analyzed, moving progressively out of the field-of-view. The blue circle in Figure 5d-i marks the position of a pole associated to the set of (002) atomic planes (Khanchandani et al., 2023).

Based on the comparison between the information gathered from TKD showing that the twin boundary is nearly perpendicular to the specimen's main axis, this APT-revealed interface cannot be the twin boundary itself, and it seems that the twin is not present in the APT data set. If the damage inside the specimen could make it simply invisible to APT analysis, it is most likely that the targeted twin boundary is no longer inside the analyzed specimen following the low-kV cleaning subsequent to TKD-mapping. This is also supported by the position of the pole in Figure 5d-i–viii that remains constant throughout the analysis, implying that there was no change in the specimen's crystallographic orientation (Gault et al., 2012).

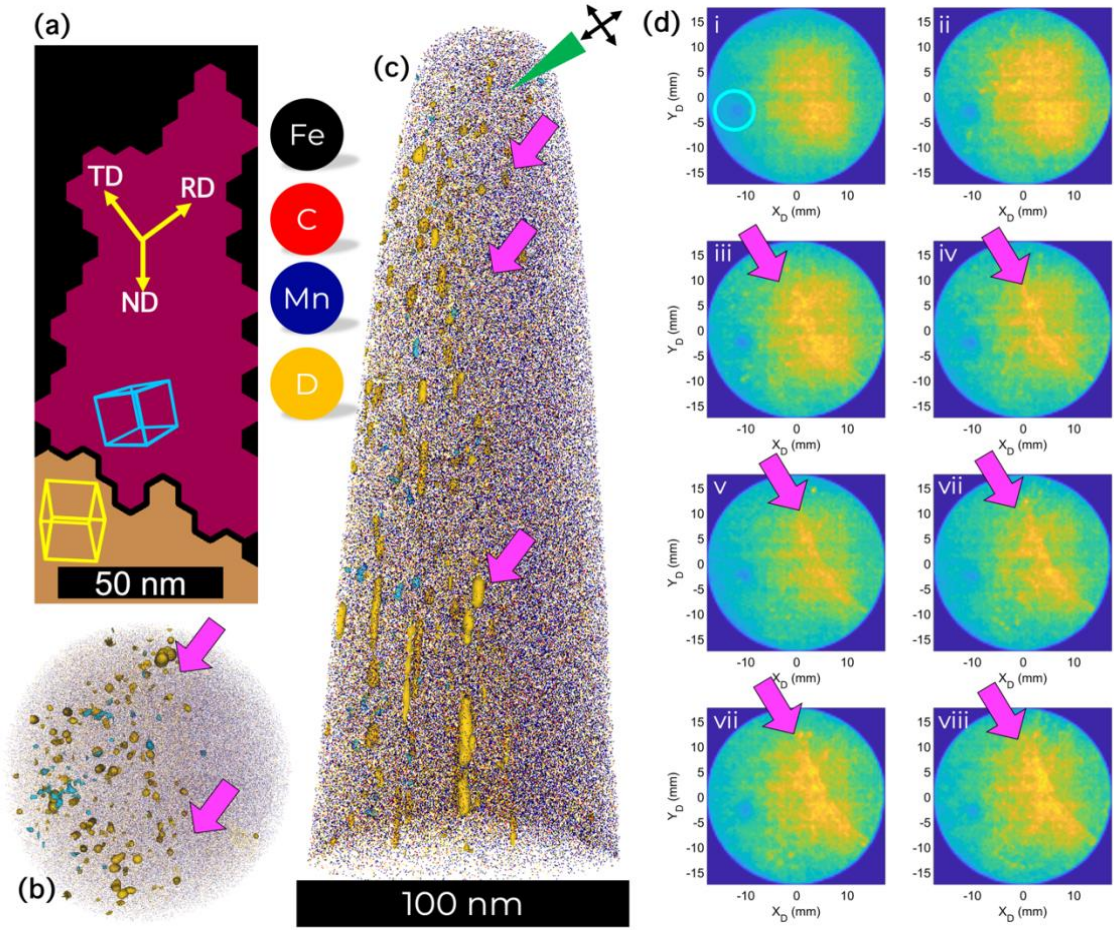

*Figure 5. (a) transmission Kikuchi diffraction unique grain map of the APT specimen. ND – normal direction in the plane, TD – transverse direction, RD – rolling direction. (b) top and (c) side view of the tomographic reconstruction from the APT analysis of the TKD-illuminated specimen. The blue isosurfaces encompass regions containing more than 1.4 at.% O and the gold isosurfaces encompass regions containing more than 3 at.% of D. The green triangle indicates the direction of the incident electron beam. (d) series of detector histograms containing each 3x10⁶ hits, and with 6x10⁶ hits in between two histograms. The blue circle marks a crystallographic pole (200). The pink arrows indicate the location of the TKD-damaged region.*

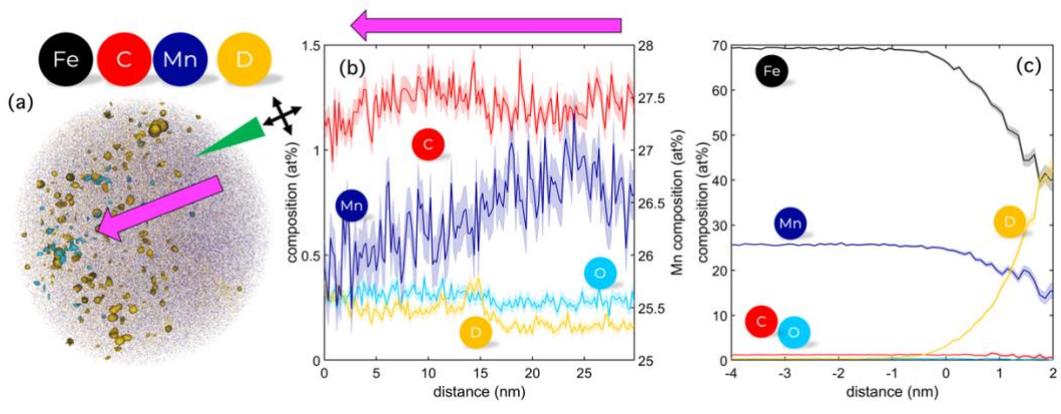

*Figure 6: (a) Top view of the 3D elemental maps with similar isosurfaces as in Figure 5 (c–d); (b) one-dimensional composition profile across the interface indicated by pink arrow in (a); Mn composition is reported on the right y-axis; (c) proximity histogram as a function of the distance to the D-isosurfaces.*

The composition of the interfacial region, Figure 6a, is analyzed using a 1D composition profile calculated in a cuboidal region-of-interest positioned perpendicular to the interface, and includes an area of 100 nm by 25 nm, Figure 6b. Deuterium is segregated up to nearly 0.4 at.%, and there is a clear enrichment towards the left side of the dataset. The D/H ratio was plotted for both datasets along with the reference data from (Khanchandani et al., 2022) in Suppl. Material Figure S2, so as to ascertain that the charging had been effective. The bulk deuterium content was 0.068 ion%, i.e. nearly three times higher than in the analysis reported in Figure 4c. O and C are not segregated to the interface, but are slightly enriched in region containing a high D-concentration, and so is H that is not plotted here since it is most likely arising from a combination of specimen preparation and residuals from the atom probe chamber (Breen et al., 2020). In this region, Mn is slightly depleted.

We calculated a single composition profile as a function of the distance to the D-isosurfaces, i.e. proximity histogram (Hellman et al., 2000), plotted in Figure 6c. Because of the difference in size between the particles highlighted by a single set of isosurfaces, this proxigram should be considered as qualitative (Martin et al., 2016). The proxigram indicates a very high concentration of D in these regions, up to 50 at% or more, which we consider to be voids or bubbles filled with D formed by the agglomeration of vacancies generated by the incoming ions. Vacancies are also likely to be further stabilized by hydrogen or deuterium (Hayward & Fu, 2013). These D-filled bubbles are similar to those reported in (Khanchandani et al., 2021b). The presence of these defects in the microstructure explains why significantly more deuterium was introduced into the specimen on which TKD mapping was performed – even if D diffuses as an interstitial, the presence of vacancies can make it more stable in the matrix (Chateau et al., 2002) thereby facilitating ingress and segregation.

The trajectory aberrations in APT around voids or bubbles lead to distortions of their shape in the reconstructed data, but also preclude precise compositional analyses due to possible intermixing of trajectories between ions originating from the matrix and inside the void or bubble (Wang et al., 2020; Dubosq et al., 2020). Linear composition profiles were calculated in cylindrical regions-of-interest positioned across four separate voids, plotted in Suppl. Material Figure S3, and the change in point density shows both the w and λ behavior observed for voids by (Wang et al., 2020). There are reports of He-filled bubbles analyzed by APT, but the analysis of the He is more challenging than D or H. We cannot conjecture as to the state of this H during the analysis. At ambient pressure, H would be liquid at the analysis temperature if it was below 21K, ie below the used 25–80K used in APT. The pressure inside the bubble is unknown, yet this would change the phase transition temperature, and how the pressure and H behave change when the void starts to be analyzed is also difficult to predict. We can still hypothesize that when the emitting surface crosses into the bubble, loosely bonded gaseous or liquid species field evaporate fast leading to a "burst" of detection and leaving a concavity at the specimen's surface that causes the severe trajectory aberrations (Wang et al., 2020; Dubosq et al., 2020). The first ionization energy of helium is approx. 24.5 eV, nearly twice that for $H_2/D_2$ that is 15.4 eV. This is reflected in the difference in the best-imaging field in field ion microscopy that is estimated to be at 45 V.nm$^{-1}$ for He and only 22 for $H_2$(Miller et al., 1996). The probability that H or $H_2$ ionizes under the electric field conditions used in our experiments (likely in the range of 25–30 Vnm$^{-1}$ for an Fe-based material in HV pulsing mode) is hence extremely high, enabling its analysis by APT.

# General discussion

Our two case studies showcase an effect of the exposure to energetic electrons during TKD mapping of APT specimens during their preparation by focused-ion beam milling. We rule out that this damage is associated to the ion beam itself, since all samples were milled with the same energy of the incoming ion beam, including for the low-energy cleaning step, and only those subjected to TKD mapping show this damage. Although this was maybe unexpected based on the general literature, we have provided some possible mechanisms, including direct damage from the incident electron beam, or, maybe more likely, the recoil of surface atoms or adsorbed light species upon electron illumination causing the equivalent damage of implanted ions.

The conditions of acceleration voltage and current used herein were similar to other reports of application of TKD to APT specimens, and this may hence be a more general problem. Future studies could aim to investigate the influence of the electron scanning strategy, the acceleration voltage and electron dose, on the amplitude of the created damage. Molecular-dynamics simulations of the damage from ion implantation in FIB milling of pure Ti indicated a contribution of thermally-activated diffusional processes on the final distribution of atoms in the sample (Chang et al., 2019). Using low temperature during TKD mapping of APT specimen was recently demonstrated in (Kim et al., 2022), but the specimen was not successfully analyzed afterwards. We ran some simulations using the package "stopping-range of ions in matter SRIM-2008.04" (Ziegler et al., 2010) for 10–30keV implantation of H and C into the alloys investigated here, using the methods introduced in Ref. (Stoller et al., 2013; Weber & Zhang, 2019) for estimating the vacancies generated due to the implantation of H and C in both materials, with an angle of incidence with respect to the target's surface of 52°, Figure 7 and Figure S4 for the TWIP steel and Ti-12Mo samples, respectively. The depth of the peak implantation is in the range of 50–120 nm for H and 10–40 nm for C, which qualitatively agrees with our observations. The structural defects, i.e. vacancies, induced by the radiation damage in the implanted region extend also beyond this peak, and are likely to diffuse following implantation.

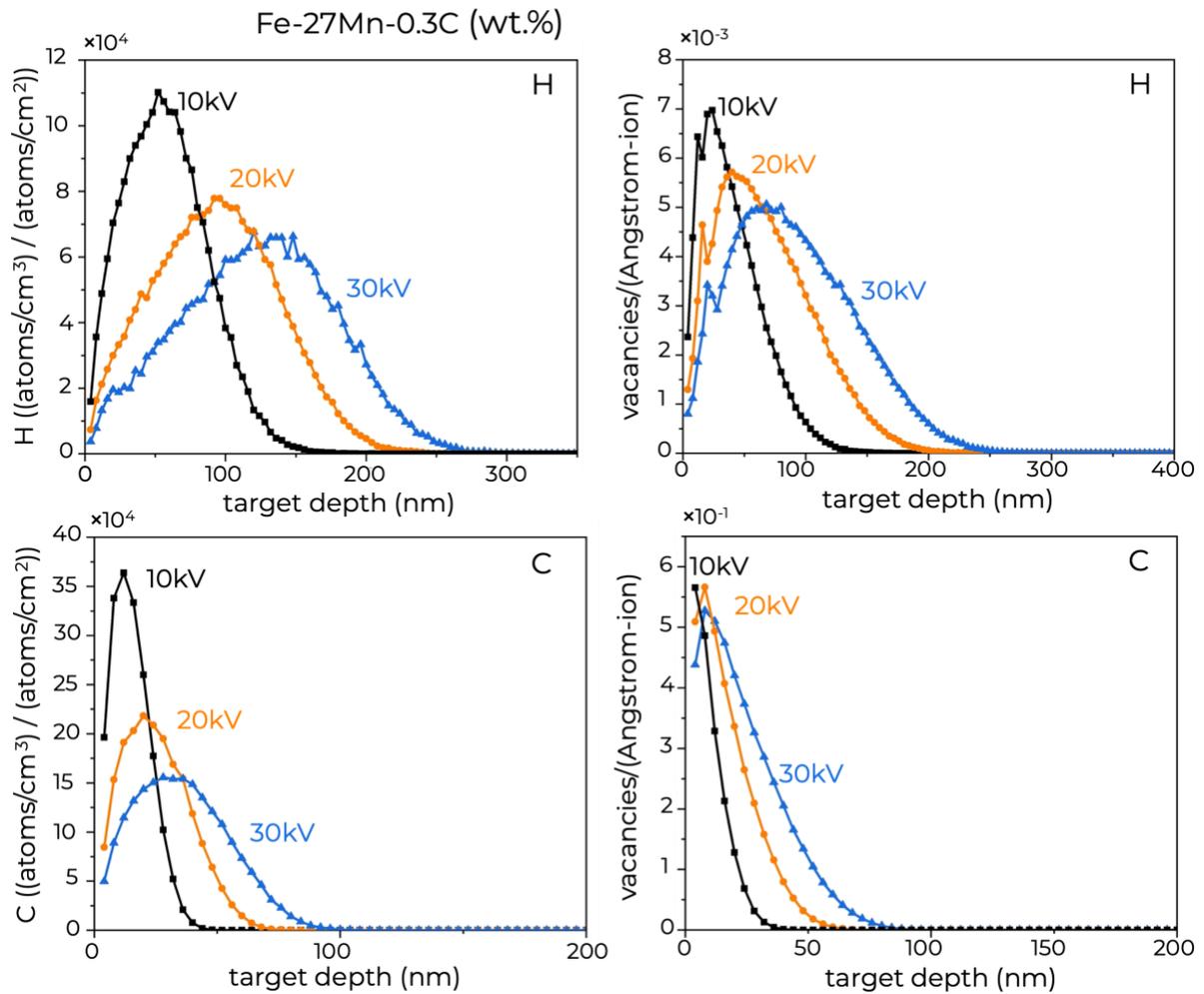

*Figure 7: SRIM-calculated implantation and vacancy depth profiles for H and C in a target with the composition of the TWIP steel analyzed herein.*

It is worth noting that if the damage inside the specimen is primarily associated to hydrogen, the composition of the specimen will make a strong difference, since the hydrogen primarily comes from the splitting of water from residual moisture that depends on the activity of the surface itself. This was reported to be particularly problematic for Ti- (Banerjee & Williams, 1983; Ding & Jones, 2011; Chang et al., 2018)and Zr-based materials (Mouton et al., 2021), in which hydrides can form. This particular sensitivity for Zr-alloys could explain the damage reported in Ref. (Gu et al., 2017). Overall, many parameters could play a role, for instance the main crystal orientation with respect to the incoming electron beam or the material's composition. In the analysis reported in Ref. (Prithiv et al., 2022) of an alloy with a close yet different composition to the Ti-12Mo, we did not observe the features associated to damage. Differences in the hydrogen ingress and behavior across different alloys had already been noted between by (Chang et al., 2018). Here again, the use of a cryogenic stage could alleviate some of the issues, as the kinetics of the water splitting could be reduced for instance, along with the slowing down of thermally-activated diffusional processes (Chang et al., 2019). Yet the possible additional condensation of e.g. water or hydrocarbons on the surfaces could be make the beam-induced damage more severe. This may depend on the time the cold sample sits in the chamber (collecting contaminants) before or during which TKD is performed, as well as the quality of the vacuum.

Finally, there are two major consequences of these artefacts related to TKD-mapping on APT which can make it arduous to distinguish real microstructural defects – for instance pre-existing vacancies or vacancy-clusters or even interfaces and grain boundaries. This has been particularly the case for the second case study in which TKD-mapping made it almost impossible to analyze the interaction of hydrogen with existing structural defects originally within the material. Concurrently, the lack of TKD mapping makes it extremely challenging to systematically study compositional effects near crystallographic defects like grain boundaries.

In addition, microstructural defects are known to alter the field evaporation characteristics (Oberdorfer et al., 2013), and there are still debates regarding if vacancies increase or decrease the evaporation field of a material (Stiller & Andren, 1982; Bowkett & Ralph, 1969), for instance. We can notice that the region in which the defect density is higher, there are differences in the composition of elements expected to be substitutionals – i.e. Mn in the Fe and Mo in Ti – from one side to the other of the specimen. Mn has a relatively lower evaporation field (Tsong, 1978; Kingham, 1982) compared to Fe, and the loss of Mn in the highly defected grain would indicate that the field evaporation proceed at a higher average electric field, agreeing with the reports by (Stiller & Andren, 1982). A similar observation applies in the Ti-12Mo analysis, there is a slightly higher average charge state of all elements on the side with a higher defect concentration. These differences in the electric field conditions during an APT analysis can explain the differences in the measured composition, associated to the difference in the evaporation field of the different species (Miller, 1981).

## Conclusions

We revealed artefacts in APT analyses that are caused by the exposure of the specimen to electrons during TKD mapping in two different metallic materials. A planar feature appears that we attribute to a peak of damage located a few tens of nanometers below the surface. These defects happen to be particularly problematic for the study of hydrogen in materials and of specific microstructural features that need to be targeted by a combination of electron microscopy and APT. Further close examination is required to understand the details of the physical origins of this artefact, its quantitative consequence on APT analyses and possible strategies to alleviate the issue.

## Acknowledgments

We thank Uwe Tezins, Christian Bross, and Andreas Sturm for their support to the FIB and APT facilities at MPIE. H.K. and B.G. acknowledge financial support from the ERC-CoG-SHINE-771602. B.G. is grateful for fruitful discussions with Dr. Aparna Saksena. T.B.B. acknowledges funding from his NSERC discovery grant.